\documentclass[preprint,showpacs,floatfix]{revtex4}

\usepackage{graphicx}
\usepackage{dcolumn}
\usepackage{bm}

\begin{document}


\title{Multipair approach to pairing in nuclei}

\author{M. Sambataro}
 \email{michelangelo.sambataro@ct.infn.it}
\affiliation{
Istituto Nazionale di Fisica Nucleare - Sezione di Catania\\
Via S. Sofia 64, I-95123 Catania, Italy}

\date{\today}

\begin{abstract}
The ground state of a general pairing Hamiltonian for a finite nuclear system is constructed as a product of collective, real, distinct pairs. These are determined sequentially via an iterative variational procedure that resorts to diagonalizations of the Hamiltonian in restricted model spaces. Different applications of the method are provided that include comparisons with exact and projected BCS results. The quantities that are examined are correlation energies, occupation numbers and pair transfer matrix elements. In a first application within the picket-fence model, the method is seen to generate the exact ground state for pairing strengths confined in a given range. Further applications of the method concern pairing in spherically symmetric mean fields and include simple exactly solvable models as well as some realistic calculations for middle-shell Sn isotopes. In the latter applications, two different ways of defining the pairs are examined: either with $J=0$ or with no well-defined angular momentum. The second choice reveals to be more effective leading, under some circumstances, to solutions that are basically exact.
\end{abstract}


\maketitle

\section{Introduction}

Since the seminal paper by Bohr, Mottelson and Pines \cite{bohr} pairing plays a crucial role in the description of both finite and infinite nuclear systems 
\cite{bohr2,ring,dean,brink}. In nuclear structure, in particular, a renewed interest in pairing has been observed in recent years following the great advances made in the experimental investigation of nuclei far from stability: pairing is essential to understand the properties of these loosely bound systems \cite{dean,brink}. 

As well known, Bohr, Mottelson and Pines transferred to nuclear physics the ideas developed by Bardeen, Cooper and Schrieffer (BCS) \cite{bardeen} to explain the electron superconductivity in metals. However, while the BCS approach turns out to be fully appropriate for macroscopic systems, being even exact in the thermodynamic limit \cite{bursill}, its application to mesoscopic systems shows some limitations owing to the fact that the BCS wave function is not an eigenstate of the number operator. In spite of that and of its being well on in years, BCS, together with the more refined Hartree-Fock-Bogoliubov (HFB) method \cite{ring}, still provides a quite common approach to pairing in nuclear structure.

Besides its inherent particle-number violation, the BCS wave function exhibits another noteworthy feature: it is formulated in terms of just one collective pair. The same feature characterizes other BCS-like approaches that have been proposed over the years to overcome the limitations of the theory (see, for instance, \cite{ring} for a review). Among these, the Projected BCS (PBCS) proposed by Blatt \cite{blatt} and Bayman
\cite{bayman} stands out for its conceptual simplicity and its effectiveness. This theory suggests a ground state wave function that is simply a condensate of one collective pair.

The scenario that the exact ground state (when available) of a pairing Hamiltonian  exhibits can be, however, very different from that suggested by PBCS. This is the case, for instance, of the so-called reduced-BCS \cite{delft} or picket-fence \cite{hirsch} model (PFM) largely used in condensed matter physics to describe the superconducting properties of ultrasmall metallic grains \cite{delft} as well as to mimic pairing in a deformed nucleus \cite{richa,gutt,duke}. The Hamiltonian of the model describes a system of fermions occupying a set of doubly degenerate equally spaced levels and interacting via a pairing force with constant strength. The corresponding eigenvalue problem can be solved exactly in a semi-analytical way \cite{richa2,richa3,richa4,richa} and one finds that, for a system of $2N$ particles, the ground state is a product of $N$ collective distinct pairs (either real or complex) irrespective of the pairing strength.

Recently Sandulescu and Bertsch \cite{sandu} carried out a detailed analysis of the validity of the BCS and PBCS (with variation after projection) approaches within the PFM. This analysis showed a very poor reliability of the BCS approximation together with, in spite of the evident dissonance between exact and approximate scenarios, a good performance of the PBCS approach. In the latter case, however, a strong dependence of the results on the size of the model space was also observed: the quality of the PBCS approximation substantially decreases with increasing the energy window around the Fermi level which fixes the model space.

A qualitative explanation for the behavior of these PBCS results can be inferred from the exact form of the ground state wave function \cite{richa2,richa3,richa4,richa}. The collective pairs defining this wave function can be of very different nature, there being pairs mainly formed by deeper bound particles as well as pairs whose basic contribution arises from particles in the upper orbitals. By enlarging the energy window around the Fermi level one actually broadens the spectrum of the pairs entering the exact wave function. Correspondingly, it becomes more and more difficult for PBCS to provide a satisfactory description of the ground state in terms of just one collective pair.

The mentioned limitations of the PBCS approach could be overcome, in principle, by resorting to a more general description of the ground state in terms of non-identical pairs. Such a description, however, is expected to be considerably more difficult than PBCS since it implies minimizing the ground state energy with respect to a much larger set of variables. Moreover, the use of pairs that are all distinct from one another does not allow the application of simple recurrence formulas to compute norms and expectation values of observables as it can be done, instead, in the PBCS case \cite{dukel}. Aiming at realizing this description by limiting as much as possible its computational cost, we have developed a simplified procedure to determine the pairs: these are constructed sequentially through an iterative sequence of diagonalizations. Each diagonalization is carried out in a space of very limited size and is meant to generate one collective pair at a time while all the others act as spectators. All pairs are by construction real. This procedure takes inspiration from a somewhat analogous iterative approach recently proposed to search for the best description of the eigenstates of a generic Hamiltonian in terms of a selected set of physically relevant configurations \cite{sambataro}. 

We will provide various applications of the method. In each case we will make a comparison with exact and PBCS results. The quantities that will be examined are correlation energies, occupation numbers and pair transfer matrix elements. The first application will concern the PFM  and will therefore simulate pairing in a deformed mean field. This will be discussed in Sec. II together with the presentation of the formalism. We will then examine some applications in spherically symmetric mean fields.      
These will include the cases of nucleons moving in a single j-shell (Sec. III A), in a double j-shell (Sec. III B) and, finally, some realistic calculations for middle-shell Sn isotopes (Sec. III C). In Sec. IV, we will summarize the results and draw some conclusions.

\section{The formalism within the PFM}

For simplicity, we will illustrate the formalism directly in the case of the PFM. 
A detailed analysis of the model can be found in the works by Richardson \cite{richa,richa2,richa4} and Richardson and Sherman \cite{richa3}. Here, we shall limit ourselves to review some of its features with special concern to the ground state.

The Hamiltonian of the model is
\begin{equation}
H=\sum^{\Omega}_{k=1}\epsilon_kN_k-g\sum^{\Omega}_{k,k'=1}P^{\dag}_kP_{k'},
\label{1}
\end{equation}
where
\begin{equation}
N_k=\sum_\sigma a^{\dag}_{k\sigma}a_{k\sigma},~~~
P^{\dag}_k=a^{\dag}_{k+}a^{\dag}_{k-},~~~ P_k=(P^{\dag}_k)^{\dag}.
\label{111}
\end{equation}
The operator $a^{\dag}_{k\sigma}$ ($a_{k\sigma}$) creates (annihilates)
a fermion in the 
single-particle state $(k,\sigma )$, where $k$ identifies one of the $\Omega$
levels of the model and $\sigma =\pm$ labels time reversed states.
These operators obey standard fermion commutation relations.
The $\Omega$ doubly degenerate levels of the model have energies
$\epsilon_k=kd$, $d$ being the level spacing.
We restrict our analysis to the case of an even number of particles $(2N)$ and exclude partial occupation of the levels, i.e. levels are
considered as either fully occupied  (two particles in time reversed states) or
empty. 

The pair product state
\begin{equation}
|\Psi\rangle =\prod^{N}_{\nu=1}B^{\dag}_{\nu}|0\rangle,~~~~~
B^{\dag}_{\nu}=\sum^{\Omega}_{k=1}\frac{1}{2\epsilon_k-e_{\nu}}P^{\dag}_k
\label{2}
\end{equation}
is an (unnormalized)
eigenstate of the Hamiltonian (\ref{1}) if the $N$ parameters $e_\nu$ 
(the so-called pair energies) are roots of
the set of $N$ coupled non-linear equations
\begin{equation}
1-\sum^{\Omega}_{k=1}\frac{g}{2\epsilon_k-e_\nu}+
\sum^{N}_{\rho=1(\rho\neq \nu)}\frac{2g}{e_\rho-e_\nu}=0.
\label{3}
\end{equation}
The eigenvalue $E^{(\Psi )}$ associated with $|\Psi\rangle$ is just the sum
of the corresponding pair energies, i.e.
\begin{equation}
E^{(\Psi )}=\sum^{N}_{\nu=1}e_\nu. 
\label{222}
\end{equation}
The pair energies $e_\nu$ can be either real or complex depending on the pairing strength $g$. In the case of the ground state and for an even number of pairs, in particular, all $e_\nu$'s turn, two by two, from real into complex-conjugate pairs with increasing the strength $g$.

The formalism of Eqs. (\ref{2})-(\ref{222}) provides a very elegant way of evaluating the ground state energy of the PFM Hamiltonian but it can only be applied within this model. More generally, the same results could be obtained by expressing the ground state as
\begin{equation}
|\Psi\rangle =\prod^{N}_{\nu=1}B^{\dag}_\nu|0\rangle,~~~~~
B^{\dag}_\nu=\sum^{\Omega}_{k=1}\beta_{k\nu}P^{\dag}_k,
\label{4}
\end{equation}
(with $\beta_{k\nu}$ complex, in general) and therefore minimizing the energy of this state with respect to the variables $\beta_{k\nu}$. With increasing the size of the system (and so of the number of variables $\beta_{k\nu}$ that should be handled simultaneously), this way of proceeding is bound to become, however, quite complicated. Owing to that and aiming at extending a description of the type (\ref{4}) to the ground state of a general pairing Hamiltonian, we have searched for an alternative (and simpler) method to determine the pairs. As a major feature, this method proposes a sequential determination of the pairs $B^{\dag}_\nu$ through an iterative  sequence of diagonalizations of the Hamiltonian in spaces of very limited size. Each diagonalization is meant to update one pair at a time while guiding the ground state energy towards its minimum. The amplitudes $\beta_{k\nu}$ are, by construction, real. 

To illustrate the method in detail,  
let $B^\dag_\nu$ $(\nu=1,...,N)$ be the pairs that at a given stage of the iterative process define the ground state $|\Psi\rangle$, Eq. (\ref{4}). We define the space
\begin{equation}
F^{(\rho)}=\Biggl\{ P^\dag_k \prod^{N}_{\nu=1(\nu\neq \rho)}B^\dag_\nu |0\rangle 
\Biggr\}
\label{5}
\end{equation}
whose states are generated by acting with all possible uncorrelated pair operators $P^\dag_k$ (\ref{111}) on a pair product state formed by all the pairs 
$B^\dag_\nu$ but the $\rho$-th one. The dimension of $F^{(\rho)}$ is therefore $\Omega$. The diagonalization of the Hamiltonian in this space generates a new ground state 
\begin{equation}
|\Psi^{(new)}\rangle =B^{\dag (new)}_\rho\prod^{N}_{\nu=1,\nu\neq \rho}B^\dag_\nu |0\rangle 
\label{6}
\end{equation}
which differs from $|\Psi\rangle$ only for the pair $B^{\dag (new)}_\rho$.
The energy of this state is by construction lower than (or, at worst, equal to) that of $|\Psi\rangle$. As a result of this operation, the pair $B^{\dag (new)}_\rho$ updates $B^{\dag}_\rho$ while all other pairs remain unchanged. Performing a series of diagonalizations of $H$ in $F^{(\rho)}$ for all possible $\rho$ values $(1\leq\rho\leq N)$ exhausts what we define a cycle of diagonalizations. At the end of a cycle all the pairs $B^\dag_\nu$ have been updated and a new cycle can then start. The iterative sequence of cycles is stopped when the difference between the ground state energies at the end of two successive cycles becomes vanishingly small.

As a first application of the method, we will study a system of $2N=16$ particles distributed over $\Omega =16$ levels. The procedure requires an initial ansatz for the pairs $B^{\dag}_\nu$ in order to start (notice, however, that only $N-1$ of these pairs are needed to generate the space $F^{(\rho)}$).
We have adopted initial pair amplitudes $\beta_{k\nu}=\delta_{k\nu}$. The state $|\Psi\rangle$ that one constructs in correspondence is nothing but the uncorrelated ground state (i.e., the state obtained by filling all levels up to the Fermi energy). We have verified, however, that different initial choices of these amplitudes do not modify significantly the results. In Fig. 1, we compare the relative errors 
$\Delta E/E=(E^{(exact)}_{corr}-E^{(appr)}_{corr})/E^{(exact)}_{corr}$ in the ground state correlation energy $E_{corr}$ that are calculated within the BCS, PBCS and present approximations. $E_{corr}$ defines the energy of the correlated ground state relative to the uncorrelated one. 

Before commenting on these results, we simply recall that the BCS approximation, for which we refer to standard textbooks (see \cite{ring}, for example), has a ground state characterized by the well known exponential form
\begin{equation}
|BCS\rangle\propto e^{B^\dag}|0\rangle=\sum_{n=0}^{\infty}\frac{(B^\dag)^n}{n!}|0\rangle,
\label{9}
\end{equation}
with the pair operator
\begin{equation}
B^{\dag}=\sum^{\Omega}_{k=1}x_kP^{\dag}_k
\label{10}
\end{equation}
being such to minimize the energy of the state (\ref{9}) under the constraint that the number of particles be conserved on average. The PBCS ground state is, instead, simply the condensate \cite{blatt,bayman}
\begin{equation}
|PBCS\rangle\propto (B^\dag)^N|0\rangle.
\label{11}
\end{equation}
The corresponding energy results from the minimization of the expectation value of the Hamiltonian in the (properly normalized) state (\ref{11})  with respect to the variables $x_k$ (variation after projection). 

The comparison of Fig. 1 refers to pairing strengths $g$ ranging in the interval (0.1, 0.9) (in units of the level spacing $d$). According to Richardson \cite{richa}, however, 
only values roughly between 0.4 and 0.7 guarantee, for the system under study, physically acceptable values of the pairing energy $P(2N)=2E(2N-1)-E(2N)-E(2N-2)$, where $E(L)$ is the ground state energy for $L$ particles.
As it is apparent from the figure, these results confirm previous conclusions \cite{richa,sandu} on the very limited reliability of the BCS approximation in finite systems (this approximation will not be further discussed in this work) and show, at the same time, a definitely better performance of the PBCS approximation whose maximum error is about $6\%$ at $g\approx 0.4$. 

For what concerns our approach, one can clearly distinguish two regions. For $g$ smaller than a critical value $g_c$ ($g_c\simeq 0.36$, indicated by the arrow in Fig. 1), this approach is able to find the exact solution. For $g\geq g_c$, instead, the error remains very small (up to three orders of magnitude smaller than the corresponding PBCS error) but nevertheless appreciably different from zero. The existence of these two regions does not come as a surprise since $g_c$ is the strength at which the Richardson pairs (\ref{2}) start being complex. This implies that, for $g\geq g_c$, the exact ground state can be represented as a pair product state only by assuming a complex form of the pairs. This complex form is not considered in our formalism which is therefore unable to find the exact solution in this region (differently from the case $g<g_c$).

We remark at this stage that, as anticipated earlier in this section, the complex Richardson pairs always occur in a complex-conjugate form, namely for every pair $B^\dag_\nu$ with a complex pair energy $e_\nu$ there always exists a pair $B^\dag_{\bar{\nu}}$ with the complex-conjugate $\bar{e}_\nu$. As shown in Ref. \cite{samba2}, the product 
$B^\dag_\nu B^\dag_{\bar{\nu}}$ can be easily rewritten as a linear combination of squares of two real pairs. Therefore, the exact PFM ground state can be equivalently formulated in terms of $N$ real pairs. In such a case, however, this state looses the simple form (\ref{2}) becoming a linear combination of pair product states. If and how a similar formalism can be extended to a general pairing Hamiltonian is an intriguing problem which deserves futher investigation.

In Figs. 2 and 3, we show the pair amplitudes $\beta_{k\nu}$ that are generated by our approach at $g=0.3$ and $g=0.7$, respectively. At $g=0.3$, these amplitudes coincide with the Richardson ones (at this strength all pairs are real). The amplitudes exhibit a peak close to one (denoting little collectivity of the pair) at the level $k=\nu$ for all pairs but the pair $\nu =8$ which appears instead visibly more collective. At $g=0.7$, the amplitudes $\beta_{k\nu}$ keep a pattern similar to that exhibited at $g=0.3$ but with a much more pronounced collectivity of the pairs. This collectivity manifestly increases moving from $\nu=1$ to $\nu=8$.

As a further test for the various approximations, we have evaluated two additional quantities: the occupation numbers $n_k=\langle\Psi |N_k|\Psi\rangle$ and the pair transfer matrix elements $t_k=|\langle\Psi(N=8)|P^\dag_k|\Psi(N=7)\rangle|$. The latter quantity has, of course, required building the ground state also for the system with $N=7$ pairs. 
In Figs. 4 and 5, we show the root mean square (rms) values $\sigma_n$ and $\sigma_t$ of the relative errors in the occupation numbers and pair transfer matrix elements, respectively, as a function of $g$. The quantities plotted are defined as
$\sigma_n=\sqrt{(\sum^\Omega_{k=1}\Delta^2_k(n))/\Omega}$  with
$\Delta_k(n)=(n^{(exact)}_k-n^{(appr)}_k)/n^{(exact)}_k$ and 
$\sigma_t=\sqrt{(\sum^\Omega_{k=1}\Delta^2_k(t))/\Omega}$  with
$\Delta_k(t)=(t^{(exact)}_k-t^{(appr)}_k)/t^{(exact)}_k$.
The behavior of these quantities is similar in the two figures and also close to that observed in Fig. 1 for the relative error in the ground state correlation energy. These new calculations confirm the very good performance of our approach by evidencing at the same time some increased difficulty for PBCS in reproducing the exact results (particularly in the pair transfer case) in regimes of very weak coupling. For completeness we also show in Figs. 6 and 7 the behaviors of the relative errors $\Delta_k(n)$ and $\Delta_k(t)$, respectively, calculated with our procedure as a function of $k$ for three different values of the pairing strength. In both cases, some peaks around the Fermi level can be observed. 

\section{Pairing in a spherically symmetric mean field}
The PFM hamiltonian (\ref{1}) studied so far schematically describes pairing in a deformed mean field. In this section, we will focus instead on pairing in a spherically symmetric mean field. The $2N$ (identical) particles of the system will be allowed to occupy a set of $2\Omega$ single-particle states labeled by the quantum numbers $n,l,j,m$ (according to the standard notation \cite{ring}). If $a^\dag_{n_il_ij_im_i}\equiv a^\dag_{im_i}$ is the operator creating a fermion in the single-particle state $im_i$ and ${\tilde{a}}^\dag_{im_i}=(-1)^{j_i-m_i}a^\dag_{i-m_i}$ is the corresponding time reversed operator, a general pairing Hamiltonian is written as
\begin{equation}
H=\sum_i\epsilon_i{\cal N}_i-\sum_{ii'}g_{ii'}L^{\dag}_iL_{i'},
\label{12}
\end{equation}
where
\begin{equation}
{\cal N}_i=\sum_{m_i}a^\dag_{im_i}a_{im_i},~~~~
L^{\dag}_i=\sum_{m_i> 0}P^\dag_{im_i},~~~~
P^\dag_{im_i}=a^\dag_{im_i}{\tilde{a}}^\dag_{im_i}.
\label{13}
\end{equation}
The operator $L^{\dag}_i$ creates a pair of particles in the state $i=n_il_ij_i$ with total angular momentum $J=0$. By further defining 
\begin{equation}
N_{im_i}=a^\dag_{im_i}a_{im_i}+a^\dag_{i-m_i}a_{i-m_i},
\label{14}
\end{equation}
the Hamiltonian (\ref{12}) becomes
\begin{equation}
H=\sum_{i,m_i>0}\epsilon_iN_{im_i}-\sum_{i,m_i>0}\sum_{i',m_{i'}>0}
g_{ii'}P^\dag_{im_i}P_{i'm_{i'}}
\end{equation}
or, using a simplified notation,
\begin{equation}
H=\sum^{\Omega}_{k=1}\epsilon_kN_k-\sum^{\Omega}_{k,k'=1}g_{kk'}P^{\dag}_kP_{k'}.
\label{15}
\end{equation}
In this expression, the index $k$ runs over the $\Omega$ levels $i, m_i>0$. Both $\epsilon_k$ and $g_{kk'}$ are independent of the projection $m$ of the angular momentum. The energies $\epsilon_k$ are therefore characterized by a $(j+1/2)$-fold degeneracy.

Due to the similarity between the Hamiltonians (\ref{1}) and (\ref{15}), the formalism of Eqs. (\ref{4})-(\ref{6}) can be applied without any change also to the present case. An important difference occurs, however, with respect to the application discussed in the previous section. As already stated, the iterative procedure requires an initial ansatz for the pair amplitudes $\beta_{k\nu}$ (Eq. (\ref{4})). In Sec. II, we have assumed $\beta_{k\nu}=\delta_{k\nu}$. This choice, however, did not significantly affect the results. This is no longer true in the cases that we are going to treat. Indeed, the mechanism of construction of the pairs is such that an initial choice of $J=0$ pairs will result in final pairs with the same angular momentum. Analogously, in correspondence with initial pairs with no well-defined angular momentum, final pairs too will not have a well-defined angular momentum, in general. As we will see in the following, however, this fact will not automatically prevent the final ground state from being a state with total angular momentum $J=0$. 

In order to examine in detail the above statements we will discuss two very simple cases of pairing, namely those of nucleons in a single $j$-shell (Sec. III A) and in a double 
$j$-shell (Sec. III B). After these illustrative examples, we will provide a realistic application of the procedure in the case of Sn isotopes (Sec. III C).

\subsection{Nucleons in a single ${\bf j}$-shell}
Pairing in a single $j$-shell represents the most elementary example of pairing in a finite nuclear system. This model allows a simple illustration of the effects that different choices of initial pairs can have on the final result in our approach. The Hamiltonian (\ref{12}) simply reduces to
\begin{equation}
H=-gL^{\dag}L
\label{16}
\end{equation}
where, for simplicity, we have omitted the one-body term and suppressed all indexes.
The (unnormalized) ground state of this Hamiltonian is represented in standard textbooks as the condensate \cite{ring} 
\begin{equation}
|\Psi_N\rangle=(L^\dag)^N|0\rangle
\label{17}
\end{equation} 
and the corresponding energy is
\begin{equation}
E_N=-gN(\Omega -N+1).
\label{18}
\end{equation}
The application of the iterative procedure of Sec. II, with initial pair amplitudes $\beta_{k\nu}=\delta_{k\nu}$ that do not guarantee any coupling to a good angular momentum, leads to the exact energy (\ref{18}) in just one iterative cycle. The wave function that is generated looks, however, very different from (\ref{17}). This is
\begin{equation}
|\Phi_N\rangle=\prod^N_{\nu=1}\Gamma^\dag_{\nu N}|0\rangle,
\label{19}
\end{equation}
with
\begin{equation}
\Gamma^\dag_{\nu N}=\sum^\Omega_{k=\nu}\gamma^{(N)}_{k\nu}P^\dag_k,~~~~~~
\gamma^{(N)}_{k\nu}=1+(N-\nu)\delta_{k\nu}.
\label{20}
\end{equation}
$|\Phi_N\rangle$ is a product of pairs that are all different from one another (notice, in particular, the index $k$ of $\Gamma^\dag_{\nu N}$ running only within the interval $(\nu, \Omega)$) and with no well-defined angular momentum. Nevertheless, this state carries a total angular momentum $J=0$: contrary to all appearances, $|\Psi_N\rangle$ and $|\Phi_N\rangle$ are actually identical. 
It is straightforward to prove analytically this identity for $N=2$ (being 
$\Gamma^\dag_{12}=L^\dag+P^\dag_1$ and $\Gamma^\dag_{22}=L^\dag-P^\dag_1$) and $N=3$  ($\Gamma^\dag_{13}=L^\dag+2P^\dag_1$, 
$\Gamma^\dag_{23}=L^\dag-P^\dag_1+P^\dag_2$, $\Gamma^\dag_{33}=L^\dag-P^\dag_1-P^\dag_2$).
This proof can be extended to any $N$, well understood that it gets more involved with increasing this number.

The case of initial J=0 pairs is a trivial one in the present example since it already provides the exact solution (\ref{17}) and the procedure only limits itself to confirm this choice. This case will be analyzed in more detail in the next application. 

\subsection{Nucleons in a double ${\bf j}$-shell}
This model assumes that nucleons are confined in two shells characterized by the same angular momentum $j$ and that they interact via a pairing force with constant strength.
The Hamiltonian of the model is therefore that of Eq. (\ref{12}) with $g_{ii'}\equiv g$ and the index $i=1,2$ labeling the two shells. We will study the case $j=11/2$ with energies $\epsilon_1 =-1$ and $\epsilon_2 =1$ (in arbitrary units) and for two different values of the strength $V={g\Omega_j}/{2\epsilon}$ (we keep the same notation of Ref. \cite{hoga}, with $\Omega_j=j+1/2$ being the half-degeneracy of the shell and $\epsilon\equiv\epsilon_2-\epsilon_1$ the difference between the single-particle energies).
In Figs. 8 and 9, upper part, we show the exact ground state correlation energy as a function of the pair number $N$ for $V=0.3$ and $V=0.7$, respectively. The lower part of the same figures shows the relative errors in this quantity that are generated by our procedure in correspondence with two different choices of the initial pairs. The line labeled with squares refers to an initial choice of random $J=0$ pairs, i.e. pairs 
$L^\dag=\sum_{i=1}^2c_iL^\dag_i$ with coefficients $c_i$ chosen randomly,
while the line labeled with triangles shows  the results for an initial choice of pairs with amplitudes $\beta_{k\nu}=\delta_{k\nu}$ (however, results do not vary significantly by assuming amplitudes $\beta_{k\nu}$ generated randomly). 

As it is apparent from these figures, results are quite different in the two cases and globally better for the ground state built from pairs with no well-defined angular momentum.
In this case, one can clearly distinguish two regions: for $N<\Omega_j$, the procedure gives rise to results that are not too far from those obtained by adopting $J=0$ pairs while, for $N\geq\Omega_j$, results turn out to be basically exact no matter the strength $V$. $N=\Omega_j$ (corresponding to the filling of the lowest shell) therefore marks a real turning point for the procedure: from this point on, the procedure becomes as effective as in the single $j$-shell case in spite of the fact that the lowest shell is only partially filled (for $N=6$ and $V=0.7$, for instance, the exact occupation numbers are $<N_1>=8.36$ and $<N_2>=3.64$).
Differently from the single $j$-shell application, however, some violations of the total angular momentum are observed (only for $N<\Omega_j$). In order to quantify these violations, we have evaluated the expectation value of the $J^2$ operator in the final ground state. The largest values found in the calculations of Figs. 8 and 9 are 0.008 at $V=0.3$ and 0.002 at $V=0.7$ always for a system with $N=5$ pairs.
It is also worthy noticing that all the pairs defining the ground state are, in this case, very far from being $J=0$ pairs: the expectation value of the $J^2$ operator for the single pairs for $N=6$ and $V=0.7$, for instance, varies from 14.0 to 51.7 .

As far as the case of initial $J=0$ pairs is concerned, the noteworthy result is that the final pairs that are generated (still with $J=0$) are all identical. In other words, in spite of being initialized with and of allowing the use of distinct $J=0$ pairs, our procedure finds the PBCS condensate as the one which guarantees the lowest energy in the model under study. 

In Figs. 10 and 11, we show the rms values of the relative errors in the occupation numbers $\langle\Psi |N_i|\Psi\rangle$ for $V=0.3$ and $V=0.7$, respectively. In Figs. 12 and 13, the corresponding quantities for the pair transfer matrix elements $|\langle\Psi (N) |L^\dag_i|\Psi (N-1)\rangle|$ are plotted. The behaviors of these quantities are consistent with the previous analysis.

We conclude this section by noticing that results qualitatively very similar to those just discussed are obtained by repeating the same calculations for a system with two different shells ($j$=11/2 and $j$=7/2).

\subsection{An application to Sn isotopes}
Pairing correlations are known to play a major role in modeling the ground state of Sn isotopes \cite{dean}. As it is common practice in the description of these isotopes, we will assume a $Z=50$, $N=50$ inert core and, dealing with systems with mass number between $A=100$ and $A=132$, we will allow neutrons to occupy the five levels $g_{7/2}$, $d_{5/2}$, $d_{3/2}$, $s_{1/2}$, and $h_{11/2}$ that are located between the magic numbers 50 and 82. Single-particle energies and pairing strengths will be the same adopted by Zelevinsky and Volya \cite{zele}. The strengths $g_{ii'}$, in particular, are derived from the interaction matrix elements $V_0(i,i')$ that result from G-matrix calculations \cite{holt}. These two quantities are related as
\begin{equation}
g_{ii'}=-V_0(i,i')/\sqrt{(j_i+1/2)(j_{i'}+1/2)}.
\label{21}
\end{equation}
The matrix elements $V_0(i,i')$ used are listed in Table I together with the single-particle energies. Being, at this stage, only interested in testing our procedure, we will not refer in the following to experimental data but rather concentrate on the comparison with exact and PBCS results.

The diagonalization  of a generic Hamiltonian in the model space just described is all but trivial for isotopes in the middle of the shell due to the large number of basis states involved \cite{holt}. In the case of the pairing Hamiltonian, however, a great simplification arises from the possibility of classifying these states within the seniority scheme \cite{raca1, raca2}. This limits the number of basis states needed to build up a ground state to a maximum value of 110 (for $^{116}$Sn) therefore making the derivation of the exact wave function straightforward \cite{zele}. In Tables II-V, we compare exact and approximate ground state correlation energies, occupation numbers and pair transfer matrix elements relative to the middle-shell $^{112-118}$Sn isotopes. As in the previous applications, we have examined both the case of initial pairs  with $J=0$ (approach A) and the case of pairs with no well-defined angular momentum (approach B). In case A, the initial pairs relative to  $^A$Sn have been assumed equal to the final ones for $^{A-2}$Sn (beginning with a simple diagonalization to find the lowest $J=0$ pair in $^{102}$Sn). In case B, we have adopted initial amplitudes $\beta_{k\nu}=\delta_{k\nu}$ as in the previous applications.

A glance at Tables II-V shows some interesting analogies with the case of nucleons in a double $j$-shell discussed in the previous section. As a general outcome, the results of approach B are always better than those of approach A in spite of some (limited) violations in the total angular momentum of the final wave function (see Tables II-V). In particular, approach B turns out to be basically exact for $^{116-118}$Sn isotopes while less effective for the lighter systems. Mass number A=114, at which this discontinuity occurs, marks a (partial) subshell closure corresponding to the filling of the levels $g_{7/2}$ and $d_{5/2}$. This closure reflects the gap in energy between these two levels and the remaining ones (see Table I). The scenario is therefore analogous to that observed in Sec. III B where, in correspondence with the partial closure of the lowest shell, one observed a drastic improvement of the results of approach B. Also in this case, the single pairs of approach B are characterized by expectation values of the $J^2$ operator that are very far from 0.
As far as approach A is concerned, instead, differently from the previous application, one finds a final ground state that is formed by pairs which are all different from one another. 
Finally, we notice the good performance of the PBCS approximation whose results, although worse than those of approaches A and B, never deviate significantly from the exact ones.

We mention that a study of pairing correlations in Sn isotopes has been recently carried out by Pillet et al. \cite{pillet3} in terms of a multiparticle-multihole configuration mixing method. This method proposes a description of the nuclear eigenstates as a linear combination of Slater determinants that include a Hartree-Fock-type state together with a (restricted) number of multiple particle-hole excitations built on this state. Both the configuration mixing coefficients and the single-particle states are determined in a self-consistent way from a variational procedure. Even though a direct comparison between the present calculations and those of Pillet et al. is difficult to make both quantitatively, due significant differences in the single-particle spaces and interactions employed, and qualitatively, due to the very different form of the two approximation schemes, we remark some common features in the two approaches: they are both variational, they preserve the particle number and they never violate the Pauli principle. 

\section{summary and conclusions}
In this paper we have searched for a description of the ground state of a general pairing Hamiltonian as a product of collective, real, distinct pairs. An iterative variational procedure has been proposed which allows a sequential determination of these pairs through the diagonalization of the Hamiltonian in spaces of very limited size. A number of applications have been carried out for both deformed and spherically symmetric systems. The procedure has proved to be effective in all these applications. Special attention has been addressed, for spherically symmetric systems, to the angular momentum of the pairs defining the ground state. We have explored both the case of $J=0$ pairs and the case of pairs with no well-defined angular momentum. In spite of generating some (limited) violations of the total angular momentum, the latter choice has revealed to be globally more effective leading to results that, under some circumstances, have been found to be basically exact even for realistic pairing Hamiltonians. 

An aspect of pairing that has attracted considerable attention in the past concerns the spatial properties of the correlations induced by this interaction. After some early studies of single pair cases like those of $^{18}$O \cite{ibarra}, $^{206}$Pb \cite{catara}, $^{210}$Pb \cite{ferreira} and, more recently,  $^{11}$Li \cite{bertsch, hagino}, the attention has been mostly concentrated on superfluid nuclei \cite{matsuo,pillet,pillet2,pastore,vinas}. The approach usually followed in these cases is the HFB (or BCS) one and the attention is focused on the spatial distribution of the abnormal density for like nucleons
$k({\vec r}_1\sigma_1,{\vec r}_2\sigma_2)=\langle \Phi|\psi({\vec r}_1\sigma_1)\psi({\vec r}_2\sigma_2)|\Phi\rangle$ \cite{ring}, where $|\Phi\rangle$ is the HFB (or BCS) ground state and $\psi({\vec r}\sigma)$ is the nucleon field operator. It is customary to regard this density as the wave function of a ``Cooper pair"  in the correlated ground state. These investigations have pointed out a small spatial distribution of this pair (2-3 fm) and its concentration in the nuclear surface \cite{matsuo,pillet,pillet2,pastore,vinas}. It has been argued \cite{dussel}, however, that this pair can only be considered as a sort of average over all the possible pairs (quite different from one another, as we have seen also in this paper) that can populate the ground state. The analysis of Ref. \cite{dussel} on $^{154}$Sm, based on the Richardson formalism, has lead to the conclusion that even the smallest of these pairs could actually be larger than the ``Cooper pair" defined above. Extending this analysis beyond the limited class of pairing Hamiltonians that can be treated in the Richardson formalism (including its possible extensions \cite{duk,duk3}) would certainly help to shed light on this debated subject. The approach presented in this paper proposes itself as a, we believe, valid tool to achieve this goal. More generally, this approach provides a new (we are not aware of similar approaches in literature) and more appropriate (with respect to standard approaches like BCS or PBCS) way to describe systems where pairs of very different nature are expected to populate the ground state. Being, however, undoubtedly more complex than these standard approaches, a greater computational effort is demanded for its applications. The impossibility, in particular, of making use of simple expressions or recurrence relations for the evaluation of norms and matrix elements of operators, as it can be done for BCS or PBCS, is an obstacle to extending the present method to systems which are instead within reach of these simpler approaches.

\begin{acknowledgments}
The author wishes to thank N. Sandulescu for his warm hospitality at NIPNE (Bucharest) and many valuable discussions on the subject of this paper.
\end{acknowledgments}

\newpage
\begin{table}
\begin{tabular}{c|c|c|c|c|c}
\colrule
& $g_{7/2}$ & $d_{5/2}$ & $d_{3/2}$ & $s_{1/2}$ & $h_{11/2}$\\
\colrule
~~~~$\epsilon_j$~~ & ~~~~$-6.121$~~~~ &  ~~~$-5.508$~~~ & ~~~$-3.749$~~~ &
 ~~~$-3.891$~~~ & ~~$-3.778$~~ \\ 
\end{tabular}
\begin{tabular}{c|ccccc}
\colrule
~$g_{7/2}$~ &  ~~$-$0.9850~~  ~~~~$-$0.5711~  ~~~~$-$0.5184~  
     ~~~$-$0.2920~  ~~~$-$1.1454~ \\ 
~$d_{5/2}$~ &  ~~~~~~~~~~  ~~~~~~~~~$-$0.7063~~  ~~~$-$0.9056~~~  
      ~$-$0.3456~~  ~~$-$0.9546~ \\ 
~$d_{3/2}$~ &  ~~~~~~~~~~~~  ~~~~~~~~~~~~~~~~~~  ~~~~~$-$0.4063~~  
      ~~$-$0.3515~~  ~~$-$0.6102~ \\ 
~$s_{1/2}$~ &  ~~~~~~~~~~~~~~  ~~~~~~~~~~~~~~~~  ~~~~~~~~~~~~~~~~~~  
      ~~$-$0.7244~~  ~~$-$0.4265~ \\ 
~$h_{11/2}$~ &  ~~~~~~~~~~~~~~  ~~~~~~~~~~~~~~~~  ~~~~~~~~~~~~~~~~~  
      ~~~~~~~~~~~~~~~~~  ~$-$1.0599~ \\ 
\colrule
\end{tabular}
\caption{Single-particle energies $\epsilon_j$ and matrix elements $V_0(j,j')$ employed in the calculations for Sn isotopes. All values are in MeV.}
\end{table}

\begin{table}
\begin{tabular}{c|c|c|c|c}
\multicolumn{5}{c}{$^{112}$Sn}\\
\colrule
& PBCS & App. A & App. B & Exact\\
\colrule
E(MeV) & $-2.8587$ &  $-2.8713$ & $-2.8954$ & $-2.9038$ \\ 
\colrule
$\Delta E/E$& 0.16 10$^{-1}$ &  0.11 10$^{-1}$ & 
0.29 10$^{-2}$& $-$ \\ 
\colrule
$\langle J^2\rangle$ & 0 & 0 & 0.45 10$^{-2}$ &  0 \\
\colrule
\multicolumn{5}{c}{$\langle N_j\rangle$}\\
\colrule
~~$j$~~& PBCS & App. A & App. B & Exact\\
\colrule
7/2 & ~~~~6.4305~~~~ & ~~~~6.4393~~~~ & ~~~~6.4602~~~~ & 
      ~~~~6.4551~~~~ \\ 
5/2 & ~~~~3.6462~~~~ & ~~~~3.6462~~~~ & ~~~~3.6338~~~~ & 
      ~~~~3.6458~~~~ \\ 
3/2 & ~~~~0.4795~~~~ & ~~~~0.4793~~~~ & ~~~~0.4771~~~~ & 
      ~~~~0.4757~~~~ \\ 
1/2 & ~~~~0.2403~~~~ & ~~~~0.2395~~~~ & ~~~~0.2370~~~~ & 
      ~~~~0.2358~~~~ \\ 
11/2 & ~~~~1.2035~~~~ & ~~~~1.1957~~~~ & ~~~~1.1919~~~~ & 
      ~~~~1.1877~~~~ \\ 
\colrule
$\sigma$ & 0.11 10$^{-1}$ &  0.84 10$^{-2}$ & 0.35 10$^{-2}$ & $-$ \\
\colrule
\multicolumn{5}{c}{$|\langle$$^{112}$Sn$|L^\dag_j|$$^{110}$Sn$\rangle|$}\\
\colrule
~~$j$~~& PBCS & App. A & App. B & Exact\\
\colrule
7/2 & ~~~~1.8997~~~~ & ~~~~1.8969~~~~ & ~~~~1.8846~~~~ & 
      ~~~~1.8870~~~~ \\ 
5/2 & ~~~~1.7035~~~~ & ~~~~1.7032~~~~ & ~~~~1.7051~~~~ & 
      ~~~~1.7040~~~~ \\ 
3/2 & ~~~~0.6588~~~~ & ~~~~0.6585~~~~ & ~~~~0.6581~~~~ & 
      ~~~~0.6569~~~~ \\ 
1/2 & ~~~~0.3312~~~~ & ~~~~0.3312~~~~ & ~~~~0.3294~~~~ & 
      ~~~~0.3287~~~~ \\ 
11/2 & ~~~~1.8225~~~~ & ~~~~1.8146~~~~ & ~~~~1.8113~~~~ & 
      ~~~~1.8089~~~~ \\ 
\colrule
$\sigma$ & 0.58 10$^{-2}$ & 0.45 10$^{-2}$ & 0.15 10$^{-2}$ & $-$ \\
\colrule
\end{tabular}
\caption{Comparison between exact and approximated ground state correlation energies, occupation numbers and pair transfer matrix elements for $^{112}$Sn. $\langle J^2 \rangle$ is the expectation value of the $J^2$ operator in the ground state. The quantities $\sigma$ are root mean square values of the relative errors. Approaches A and B are described in the text.}
\end{table}

\begin{table}
\begin{tabular}{c|c|c|c|c}
\multicolumn{5}{c}{$^{114}$Sn}\\
\colrule
& PBCS & App. A & App. B & Exact\\
\colrule
E(MeV) & $-2.5237$ &  $-2.5427$ & $-2.6002$ & $-2.6011$ \\ 
\colrule
$\Delta E/E$& 0.30 10$^{-1}$ &  0.22 10$^{-1}$ & 
0.34 10$^{-3}$& $-$ \\ 
\colrule
$\langle J^2\rangle$ & 0 & 0 & 0.18 10$^{-3}$ &  0 \\
\colrule
\multicolumn{5}{c}{$\langle N_j\rangle$}\\
\colrule
~~$j$~~& PBCS & App. A & App. B & Exact\\
\colrule
7/2 & ~~~~6.9182~~~~ & ~~~~6.9438~~~~ & ~~~~6.9568~~~~ & 
      ~~~~6.9562~~~~ \\ 
5/2 & ~~~~4.3909~~~~ & ~~~~4.4218~~~~ & ~~~~4.4638~~~~ & 
      ~~~~4.4630~~~~ \\ 
3/2 & ~~~~0.6478~~~~ & ~~~~0.6379~~~~ & ~~~~0.6271~~~~ & 
      ~~~~ 0.6265~~~~ \\ 
1/2 & ~~~~0.3743~~~~ & ~~~~0.3674~~~~ & ~~~~0.3536~~~~ & 
      ~~~~0.3558~~~~ \\ 
11/2 & ~~~~1.6687~~~~ & ~~~~1.6290~~~~ & ~~~~1.5988~~~~ & 
      ~~~~1.5985~~~~ \\ 
\colrule
$\sigma$ & 0.35 10$^{-1}$ &  0.19 10$^{-1}$ & 0.27 10$^{-2}$ & $-$ \\
\colrule
\multicolumn{5}{c}{$|\langle$$^{114}$Sn$|L^\dag_j|$$^{112}$Sn$\rangle|$}\\
\colrule
~~$j$~~& PBCS & App. A & App. B & Exact\\
\colrule
7/2 & ~~~~1.6453~~~~ & ~~~~1.6370~~~~ & ~~~~1.6126~~~~ & 
      ~~~~1.6197~~~~ \\ 
5/2 & ~~~~1.6056~~~~ & ~~~~1.6087~~~~ & ~~~~1.6153~~~~ & 
      ~~~~1.6107~~~~ \\ 
3/2 & ~~~~0.7542~~~~ & ~~~~0.7476~~~~ & ~~~~0.7415~~~~ & 
      ~~~~0.7411~~~~ \\ 
1/2 & ~~~~0.4053~~~~ & ~~~~0.4019~~~~ & ~~~~0.3940~~~~ & 
      ~~~~0.3958~~~~ \\ 
11/2 & ~~~~2.1197~~~~ & ~~~~2.0898~~~~ & ~~~~2.0667~~~~ & 
      ~~~~2.0687~~~~ \\ 
\colrule
$\sigma$ & 0.19 10$^{-1}$ & 0.10 10$^{-1}$ & 0.31 10$^{-2}$ & $-$ \\
\colrule
\end{tabular}
\caption{As in Table II, for $^{114}$Sn.}
\end{table}

\begin{table}
\begin{tabular}{c|c|c|c|c}
\multicolumn{5}{c}{$^{116}$Sn}\\
\colrule
& PBCS & App. A & App. B & Exact\\
\colrule
E(MeV) & $-3.5703$ &  $-3.5925$ & $-3.6185$ & $-3.6185$ \\ 
\colrule
$\Delta E/E$& 0.13 10$^{-1}$ &  0.72 10$^{-2}$ & 
0.16 10$^{-4}$& $-$ \\ 
\colrule
$\langle J^2\rangle$ & 0 & 0 & 0.14 10$^{-4}$ &  0 \\
\colrule
\multicolumn{5}{c}{$\langle N_j\rangle$}\\
\colrule
~~$j$~~& PBCS & App. A & App. B & Exact\\
\colrule
7/2 & ~~~~7.1334~~~~ & ~~~~7.1407~~~~ & ~~~~7.1413~~~~ & 
      ~~~~7.1413~~~~ \\ 
5/2 & ~~~~4.7479~~~~ & ~~~~4.7618~~~~ & ~~~~4.7697~~~~ & 
      ~~~~4.7697~~~~ \\ 
3/2 & ~~~~0.9391~~~~ & ~~~~0.9295~~~~ & ~~~~0.9280~~~~ & 
      ~~~~0.9280~~~~ \\ 
1/2 & ~~~~0.6332~~~~ & ~~~~0.6557~~~~ & ~~~~0.6463~~~~ & 
      ~~~~0.6463~~~~ \\ 
11/2 & ~~~~2.5465~~~~ & ~~~~2.5124~~~~ & ~~~~2.5146~~~~ & 
      ~~~~2.5147~~~~ \\ 
\colrule
$\sigma$ & 0.12 10$^{-1}$ &  0.66 10$^{-2}$ & 0.15 10$^{-4}$ & $-$ \\
\colrule
\multicolumn{5}{c}{$|\langle$$^{116}$Sn$|L^\dag_j|$$^{114}$Sn$\rangle|$}\\
\colrule
~~$j$~~& PBCS & App. A & App. B & Exact\\
\colrule
7/2 & ~~~~1.3846~~~~ & ~~~~1.3628~~~~ & ~~~~1.3585~~~~ & 
      ~~~~1.3592~~~~ \\ 
5/2 & ~~~~1.3775~~~~ & ~~~~1.3630~~~~ & ~~~~1.3489~~~~ & 
      ~~~~1.3494~~~~ \\ 
3/2 & ~~~~0.8840~~~~ & ~~~~0.8773~~~~ & ~~~~0.8720~~~~ & 
      ~~~~0.8722~~~~ \\ 
1/2 & ~~~~0.5062~~~~ & ~~~~0.5150~~~~ & ~~~~0.5140~~~~ & 
      ~~~~0.5138~~~~ \\ 
11/2 & ~~~~2.5558~~~~ & ~~~~2.5343~~~~ & ~~~~2.5249~~~~ & 
      ~~~~2.5256~~~~ \\ 
\colrule
$\sigma$ & 0.16 10$^{-1}$ & 0.57 10$^{-2}$ & 0.38 10$^{-3}$ & $-$ \\
\colrule
\end{tabular}
\caption{As in Table II, for $^{116}$Sn.}
\end{table}

\begin{table}
\begin{tabular}{c|c|c|c|c}
\multicolumn{5}{c}{$^{118}$Sn}\\
\colrule
& PBCS & App. A & App. B & Exact\\
\colrule
E(MeV) & $-4.0749$ &  $-4.0941$ & $-4.1052$ & $-4.1053$ \\ 
\colrule
$\Delta E/E$& 0.74 10$^{-2}$ &  0.27 10$^{-2}$ & 
0.56 10$^{-5}$& $-$ \\ 
\colrule
$\langle J^2\rangle$ & 0 & 0 & 0.36 10$^{-5}$ &  0 \\
\colrule
\multicolumn{5}{c}{$\langle N_j\rangle$}\\
\colrule
~~$j$~~& PBCS & App. A & App. B & Exact\\
\colrule
7/2 & ~~~~7.2711~~~~ & ~~~~7.2732~~~~ & ~~~~7.2727~~~~ & 
      ~~~~7.2727~~~~ \\ 
5/2 & ~~~~4.9665~~~~ & ~~~~4.9730~~~~ & ~~~~4.9743~~~~ & 
      ~~~~4.9743~~~~ \\ 
3/2 & ~~~~1.2624~~~~ & ~~~~1.2535~~~~ & ~~~~1.2526~~~~ & 
      ~~~~1.2526~~~~ \\ 
1/2 & ~~~~0.9034~~~~ & ~~~~0.9230~~~~ & ~~~~0.9171~~~~ & 
      ~~~~0.9171~~~~ \\ 
11/2 & ~~~~3.5966~~~~ & ~~~~3.5774~~~~ & ~~~~3.5833~~~~ & 
      ~~~~3.5833~~~~ \\ 
\colrule
$\sigma$ & 0.77 10$^{-2}$ &  0.30 10$^{-2}$ & 0.80 10$^{-5}$ & $-$ \\
\colrule
\multicolumn{5}{c}{$|\langle$$^{118}$Sn$|L^\dag_j|$$^{116}$Sn$\rangle|$}\\
\colrule
~~$j$~~& PBCS & App. A & App. B & Exact\\
\colrule
7/2 & ~~~~1.2513~~~~ & ~~~~1.2451~~~~ & ~~~~1.2466~~~~ & 
      ~~~~1.2466~~~~ \\ 
5/2 & ~~~~1.2430~~~~ & ~~~~1.2357~~~~ & ~~~~1.2336~~~~ & 
      ~~~~1.2336~~~~ \\ 
3/2 & ~~~~0.9792~~~~ & ~~~~0.9743~~~~ & ~~~~0.9719~~~~ & 
      ~~~~0.9720~~~~ \\ 
1/2 & ~~~~0.5541~~~~ & ~~~~0.5553~~~~ & ~~~~0.5563~~~~ & 
      ~~~~0.5563~~~~ \\ 
11/2 & ~~~~2.9049~~~~ & ~~~~2.8975~~~~ & ~~~~2.8951~~~~ & 
      ~~~~2.8951~~~~ \\ 
\colrule
$\sigma$ & 0.56 10$^{-2}$ & 0.16 10$^{-2}$ & 0.21 10$^{-4}$ & $-$ \\
\colrule
\end{tabular}
\caption{As in Table II, for $^{118}$Sn.}
\end{table}

\newpage
\begin{figure}
\includegraphics{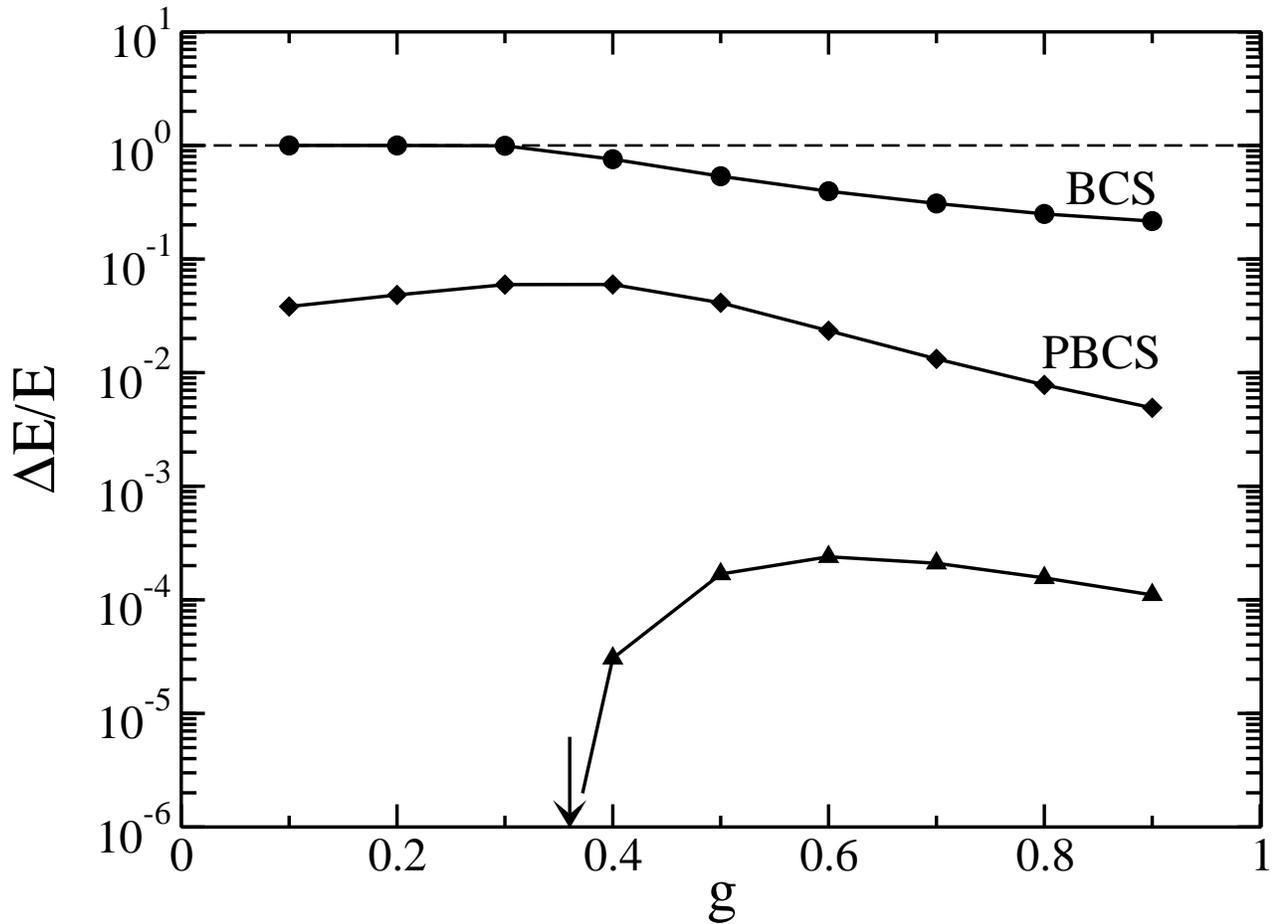}
\mbox{}\\[19cm]
\caption{Relative errors in the ground state correlation energy calculated for a system with $2N=\Omega =16$ particles within BCS, PBCS and the present approach (line labeled with triangles) as a function of the pairing strength $g$. The arrow indicates the critical value $g_c$ at which the Richardson pairs start being complex (see text). $g$ is in units of the level spacing $d$.}
\end{figure}

\begin{figure}
\includegraphics{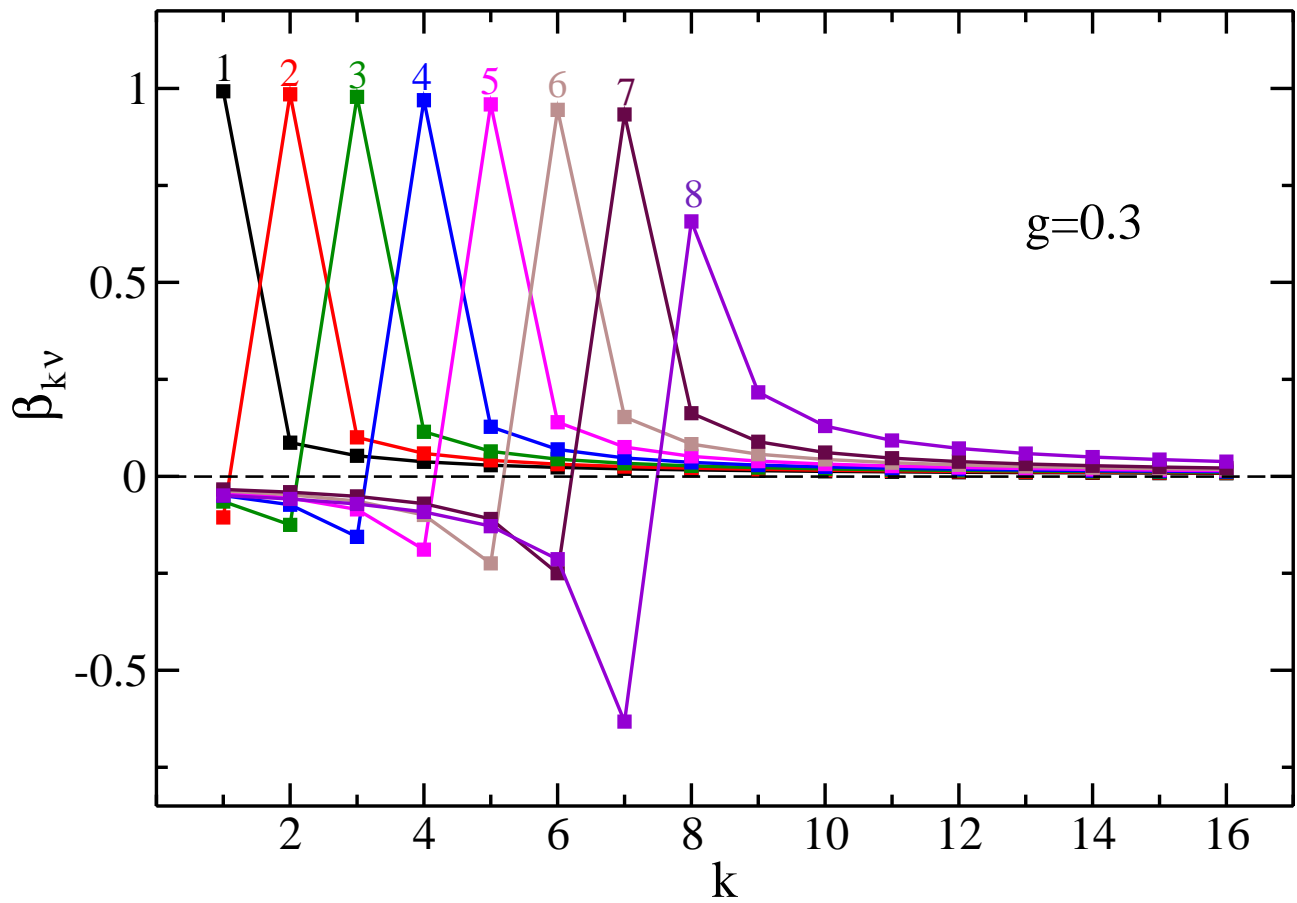}
\mbox{}\\[19cm]
\caption{(Color online) Pair amplitudes $\beta_{k\nu}$ (Eq. (\ref{4})) calculated within the present approach for a system with $2N=\Omega =16$ particles  and a pairing strength $g=0.3$ (in units of $d$). Each line shows the amplitudes $\beta_{k\nu}$ relative to the pair $\nu$ indicated on top of it. In the figure, the amplitudes relative to each pair $\nu$ have been assigned an arbitrary overall phase such that $\beta_{k\nu}>0$ at $k=16$. The normalization is such that $\sum_{k}\beta^2_{k\nu}=1$.}
\end{figure}

\begin{figure}
\includegraphics{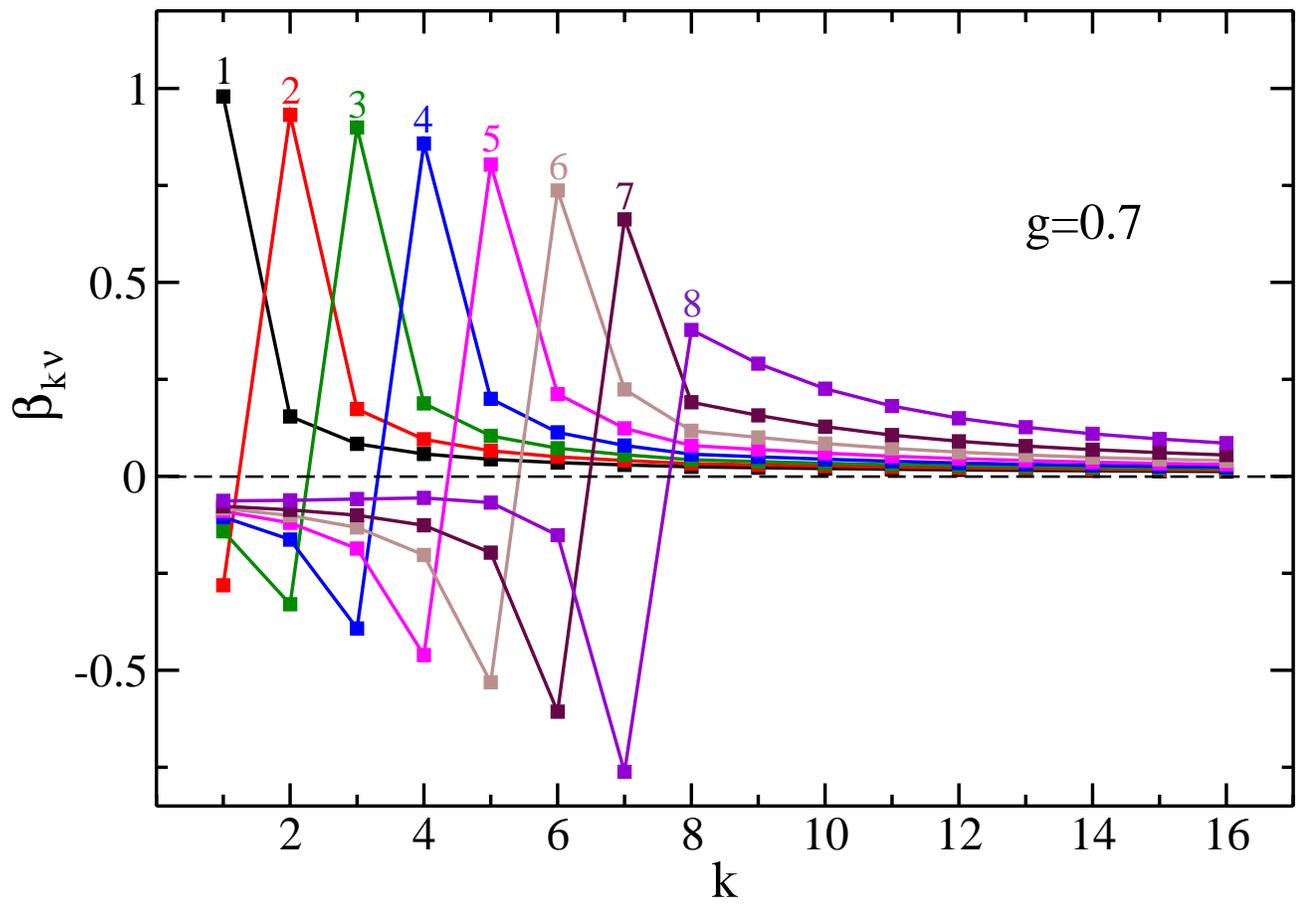}
\mbox{}\\[19cm]
\caption{(Color online) As in Fig. 2, but for $g=0.7$.}
\end{figure}

\begin{figure}
\includegraphics{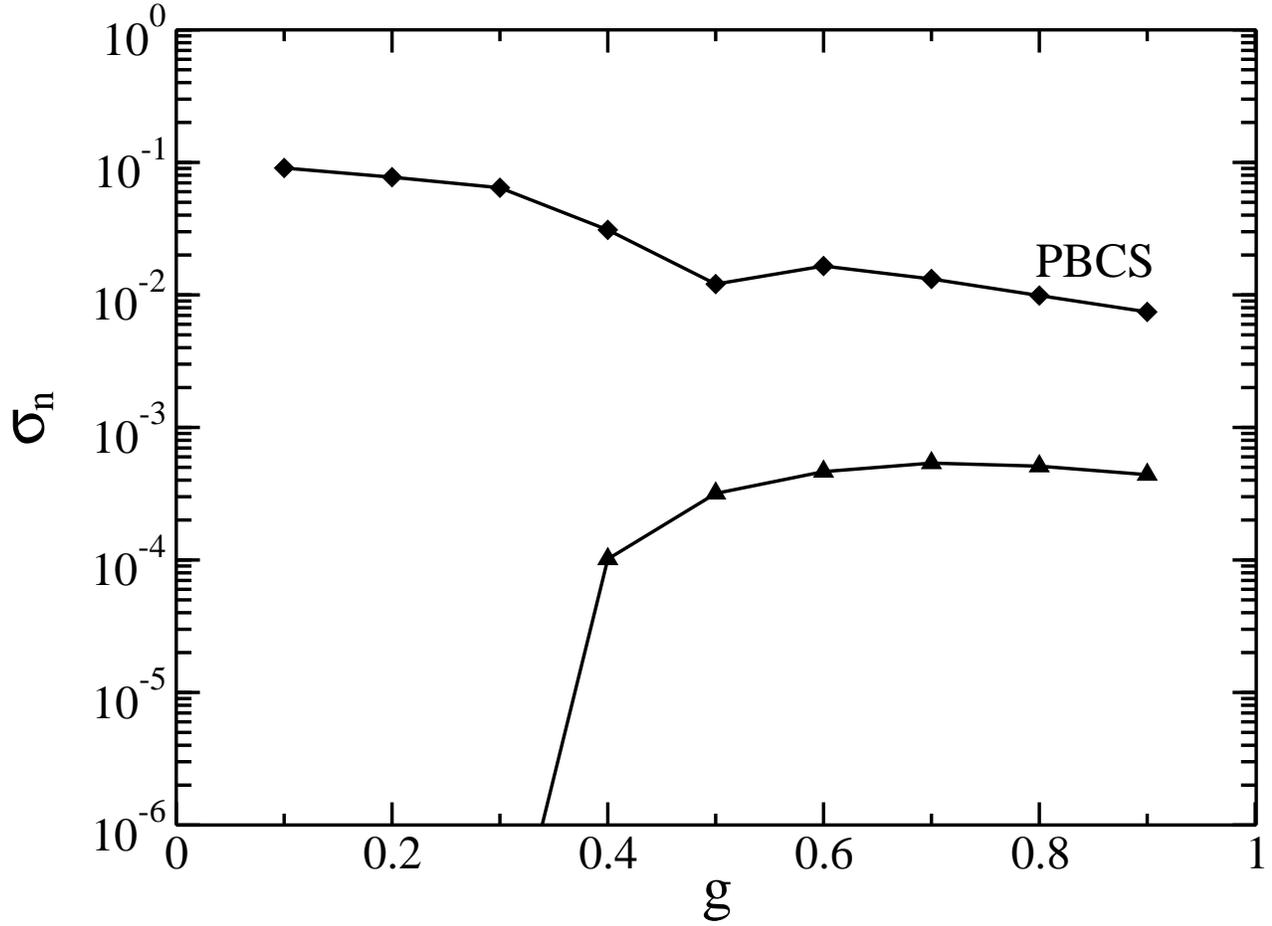}
\mbox{}\\[19cm]
\caption{Root mean square values of the relative errors in the occupation numbers calculated  for a system with $2N=\Omega =16$ particles within PBCS and the present approach (line labeled with triangles) as a function of the pairing strength $g$ (in units of $d$).}
\end{figure}

\begin{figure}
\includegraphics{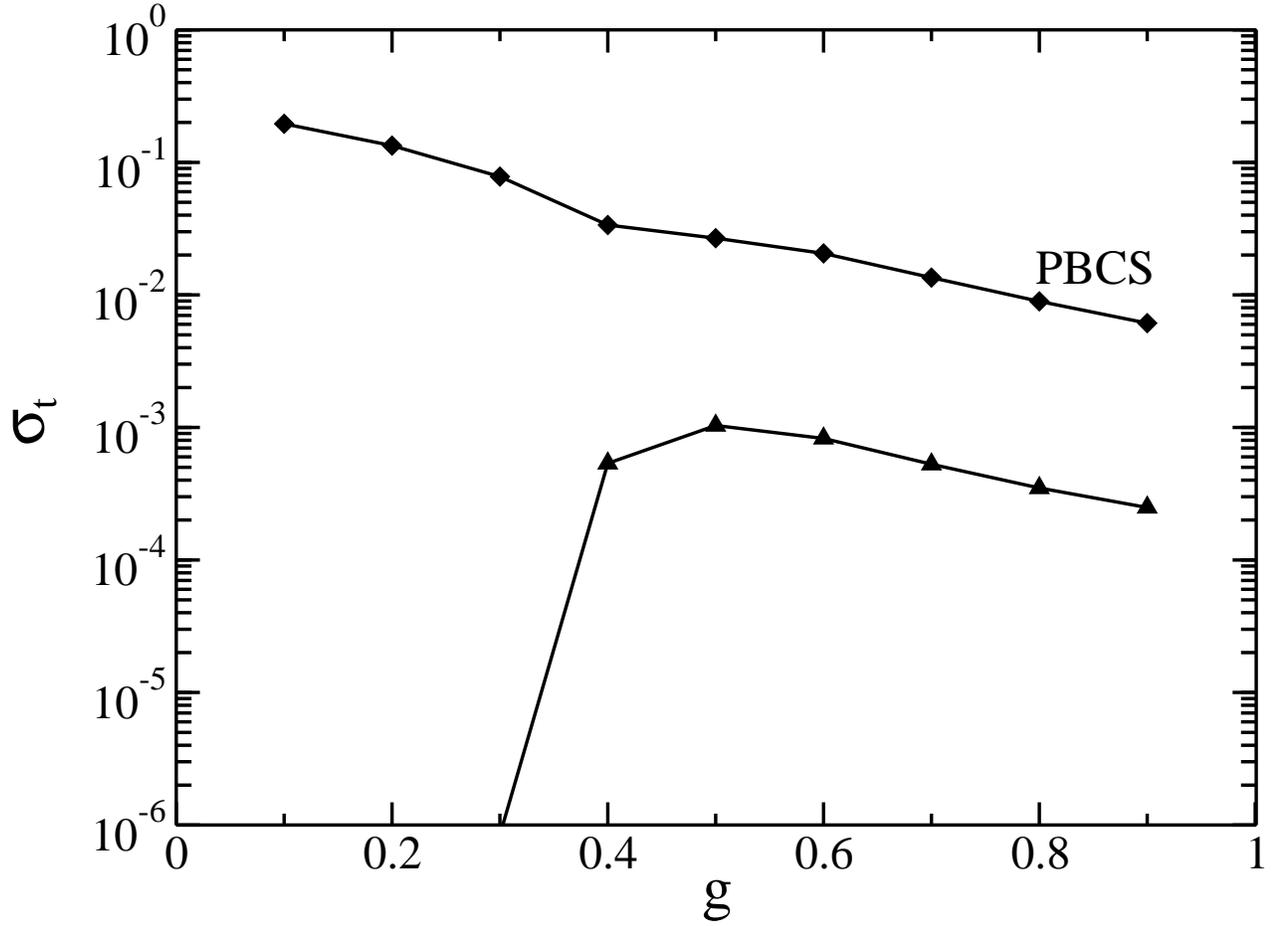}
\mbox{}\\[19cm]
\caption{Root mean square values of the relative errors in the pair transfer matrix elements calculated  for a system with $2N=\Omega =16$ particles within PBCS and the present approach (line labeled with triangles) as a function of the pairing strength $g$ (in units of $d$).}
\end{figure}

\begin{figure}
\includegraphics{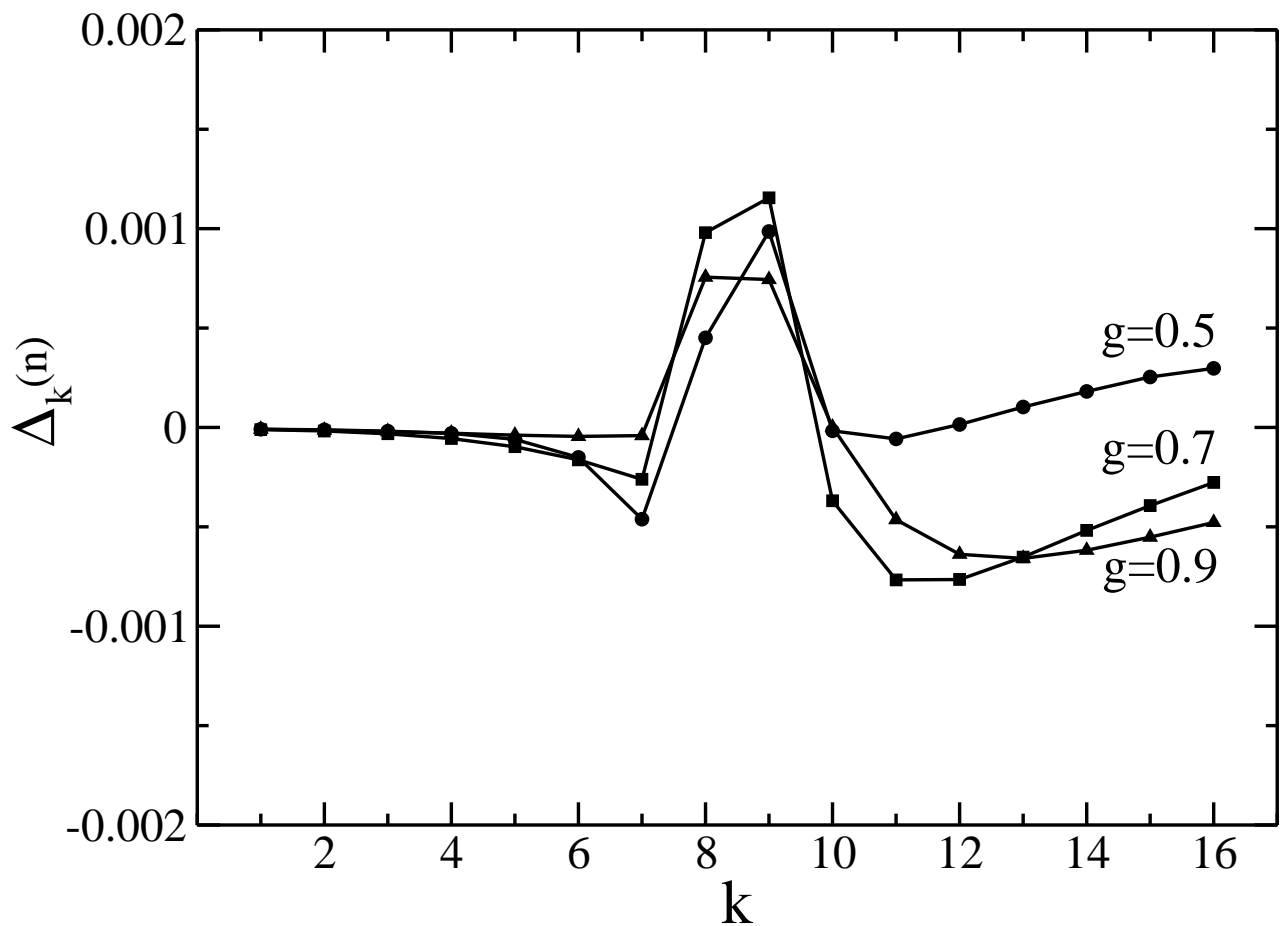}
\mbox{}\\[19cm]
\caption{Relative errors in the occupation numbers $n_k$ calculated with our procedure as a function of $k$ for a system with $2N=\Omega =16$ particles. The lines refer to three different values of the pairing strength.}
 \end{figure}

\begin{figure}
\includegraphics{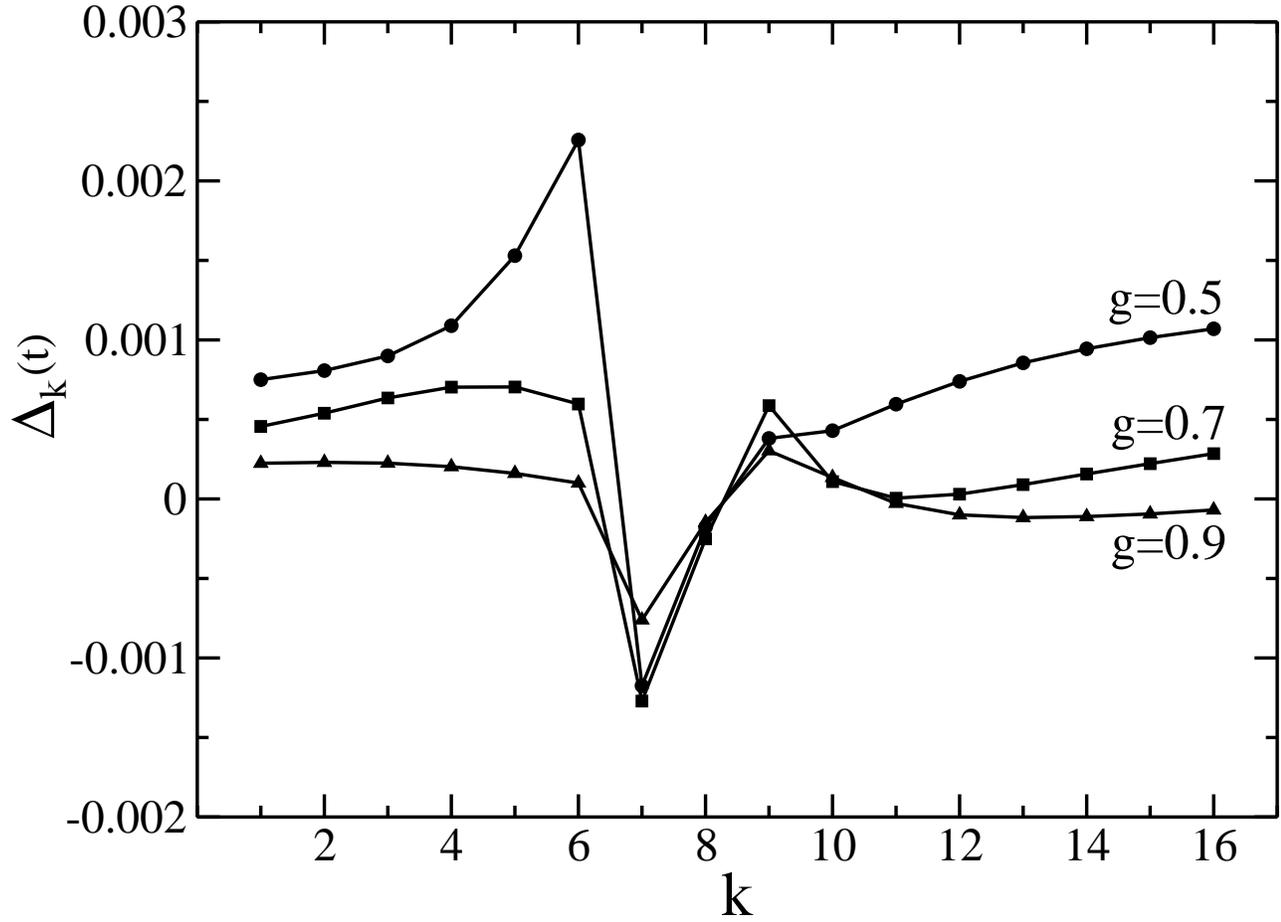}
\mbox{}\\[19cm]
\caption{Relative errors in the pair transfer matrix elements $t_k$ calculated with our procedure as a function of $k$ for a system with $2N=\Omega =16$ particles. The lines refer to three different values of the pairing strength.}
\end{figure}

\begin{figure}
\includegraphics{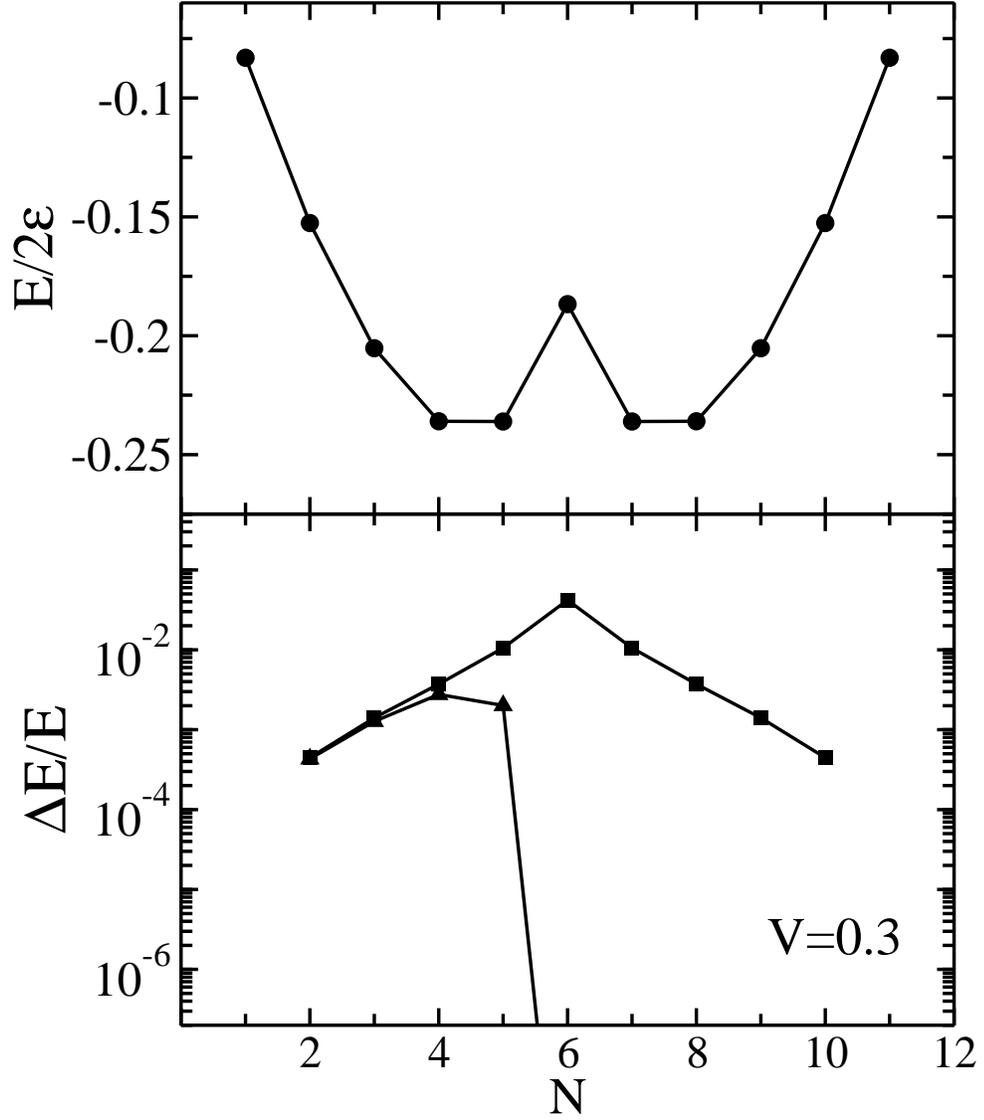}
\mbox{}\\[19cm]
\caption{Upper part:  exact ground state correlation energy for $2N$ particles in two $j=11/2$ shells and $V=0.3$. The energy is expressed in units of $2\epsilon$. Lower part: relative errors in the ground state correlation energy calculated with the present procedure
in correspondence with two different choices of the initial pairs. 
Squares refer to  $J=0$ pairs while triangles to pairs with no well-defined angular momentum (see text for details).}
\end{figure}

\begin{figure}
\includegraphics{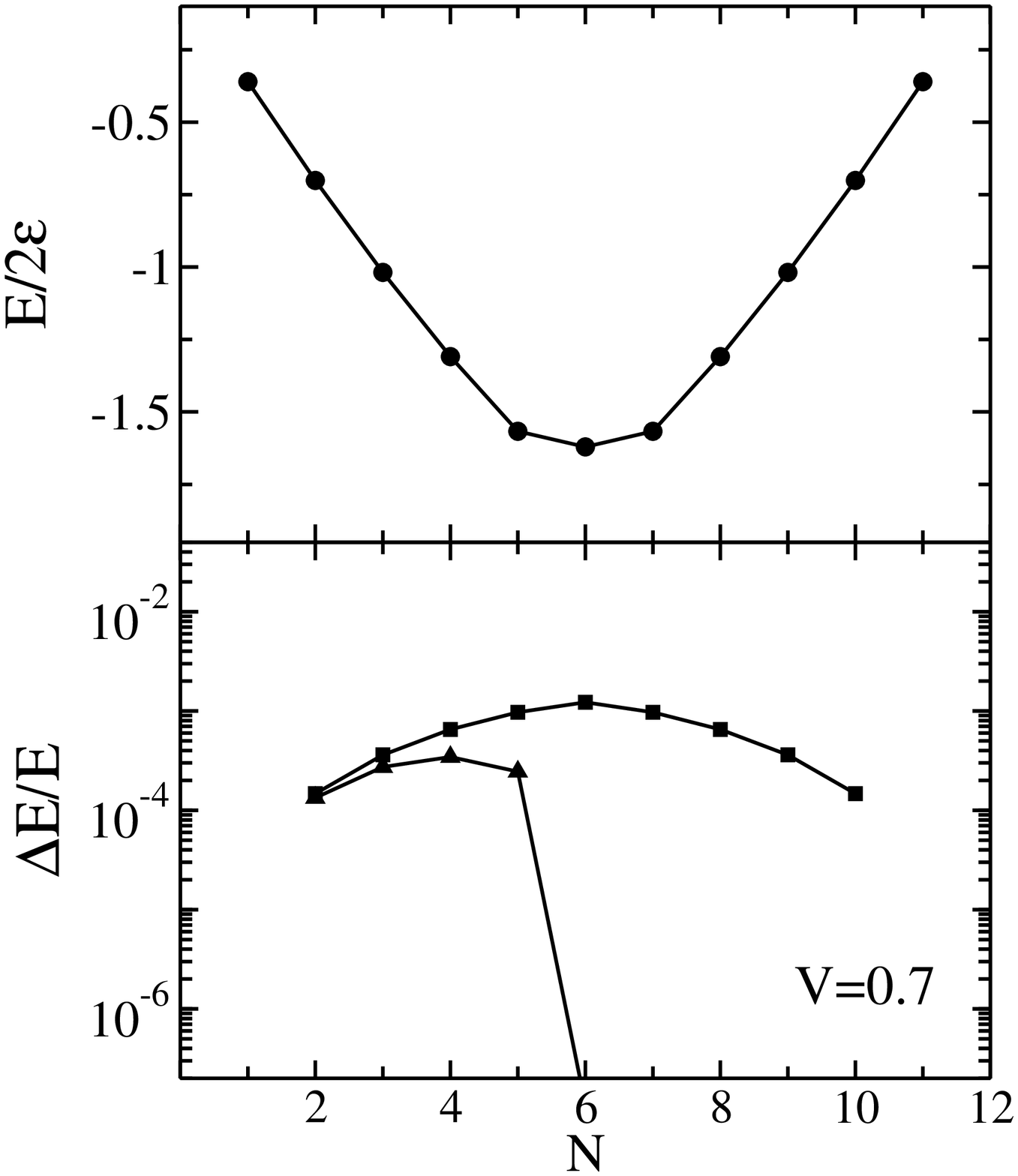}
\mbox{}\\[19cm]
\caption{As in Fig. 6, for  $V=0.7$.}
\end{figure}

\begin{figure}
\includegraphics{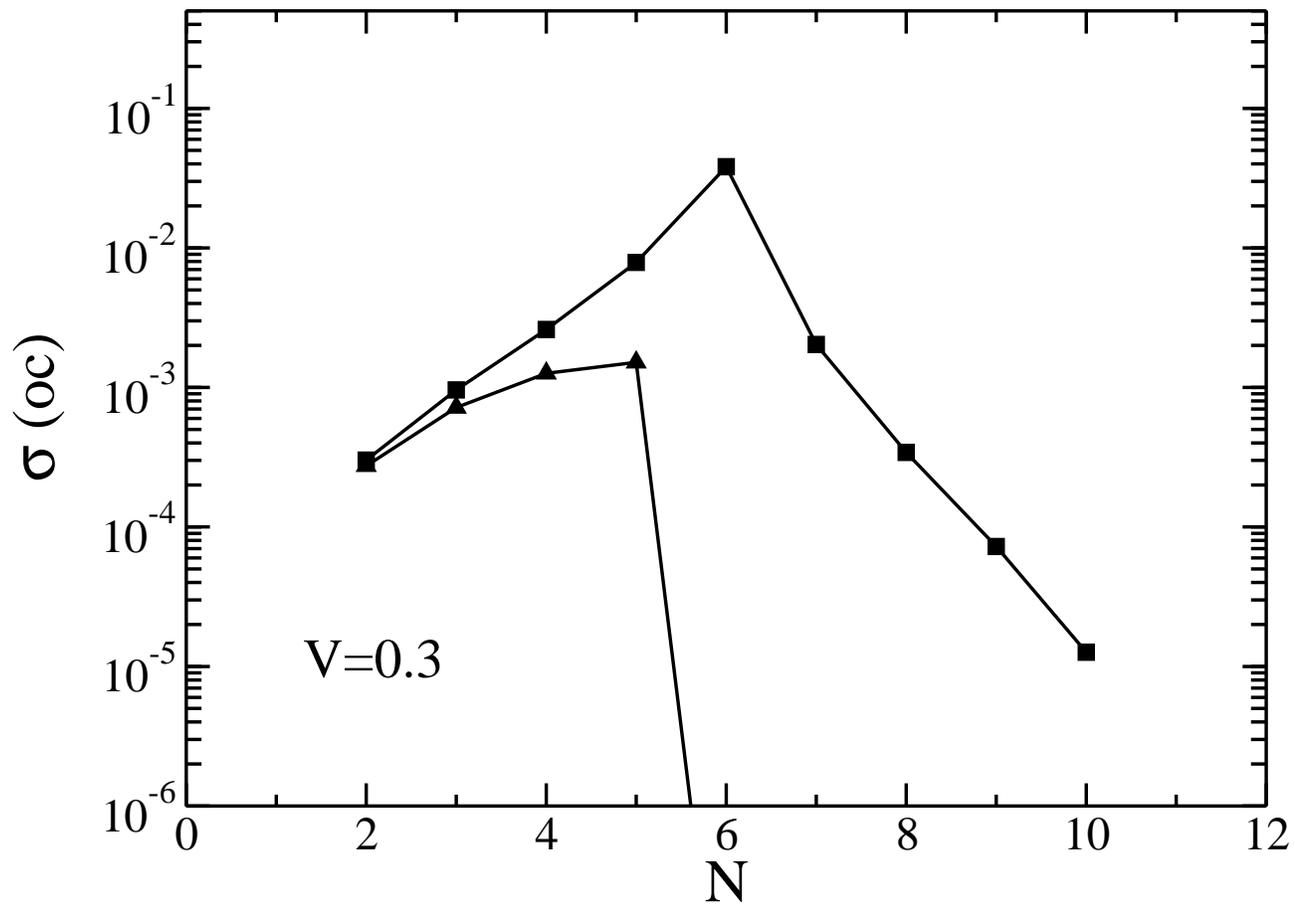}
\mbox{}\\[19cm]
\caption{Root mean square values of the relative errors in the occupation numbers calculated with the present procedure for $2N$ particles in two $j=11/2$ shells and $V=0.3$. The two lines refer to different choices of the initial pairs: $J=0$ pairs (squares) and pairs with no well-defined angular momentum (triangles). See text for details.} 
\end{figure}

\begin{figure}
\includegraphics{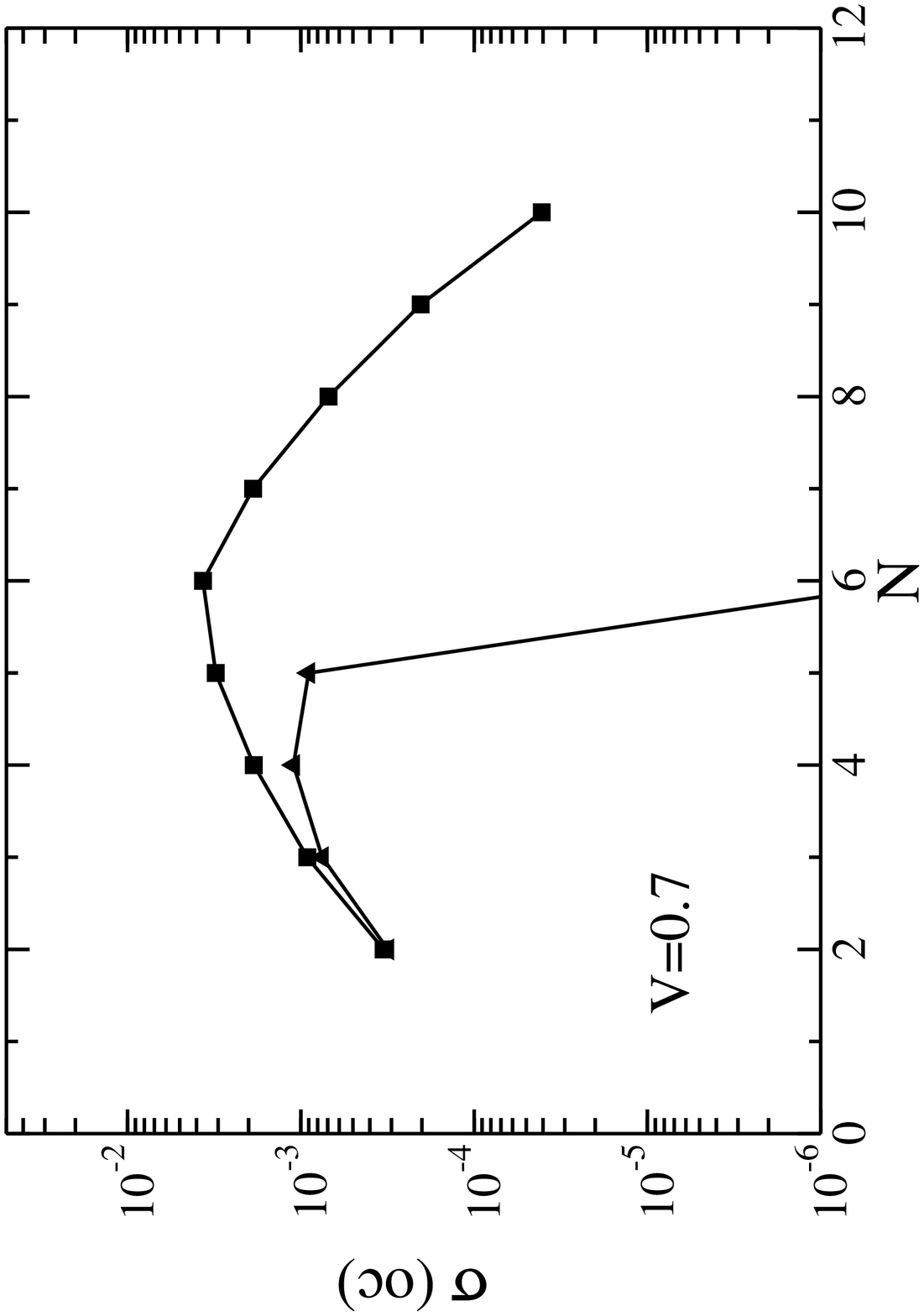}
\mbox{}\\[19cm]
\caption{As in Fig. 8, for $V=0.7$.}
\end{figure}

\begin{figure}
\includegraphics{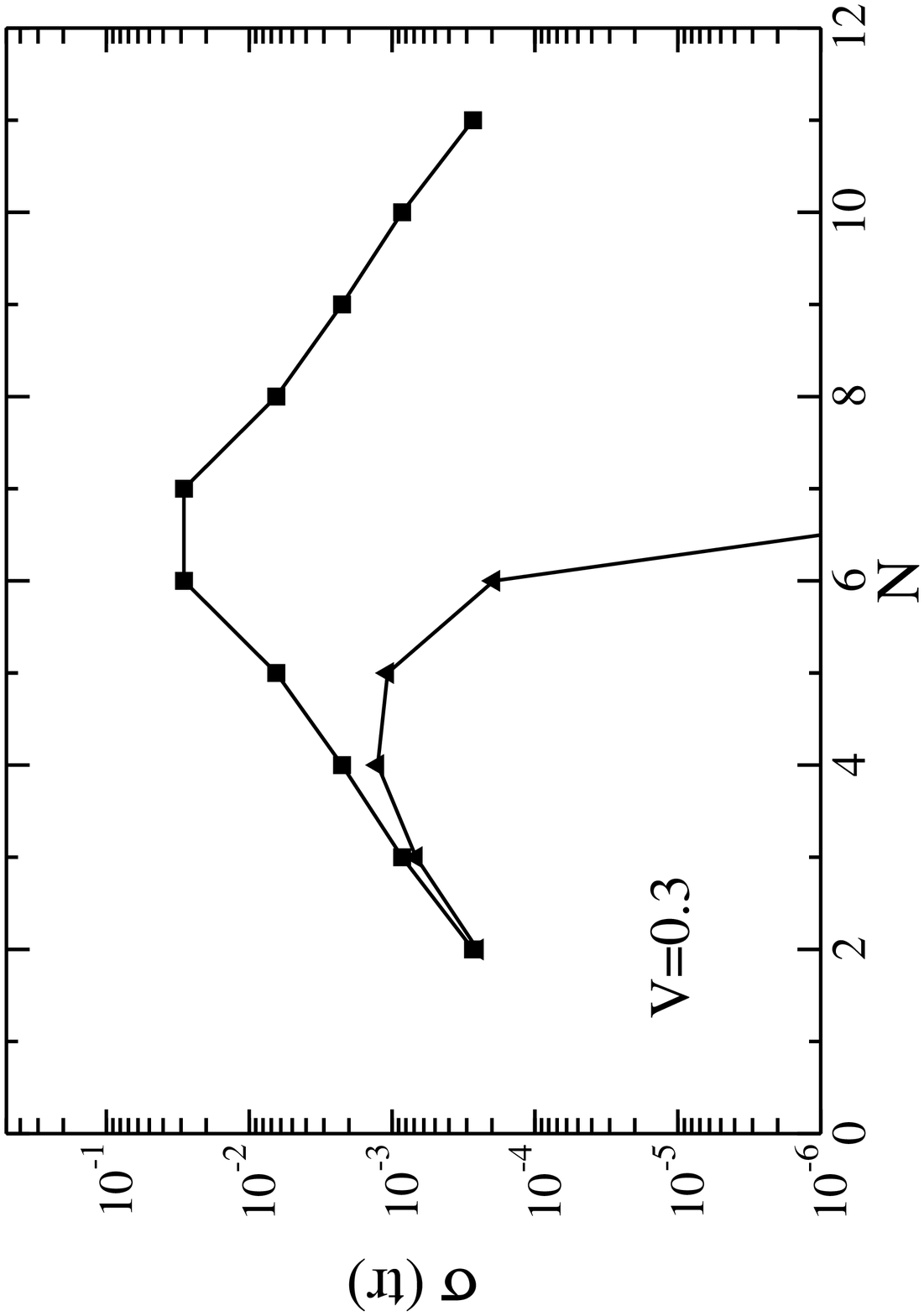}
\mbox{}\\[19cm]
\caption{Root mean square values of the relative errors in the pair transfer matrix elements $|\langle\Psi (N) |L^\dag_i|\Psi (N-1)\rangle|$ calculated with the present procedure for $2N$ particles in two $j=11/2$ shells and $V=0.3$. The two lines refer to different choices of the initial pairs: $J=0$ pairs (squares) and pairs with no well-defined angular momentum (triangles). See text for details.} 
\end{figure}

\begin{figure}
\includegraphics{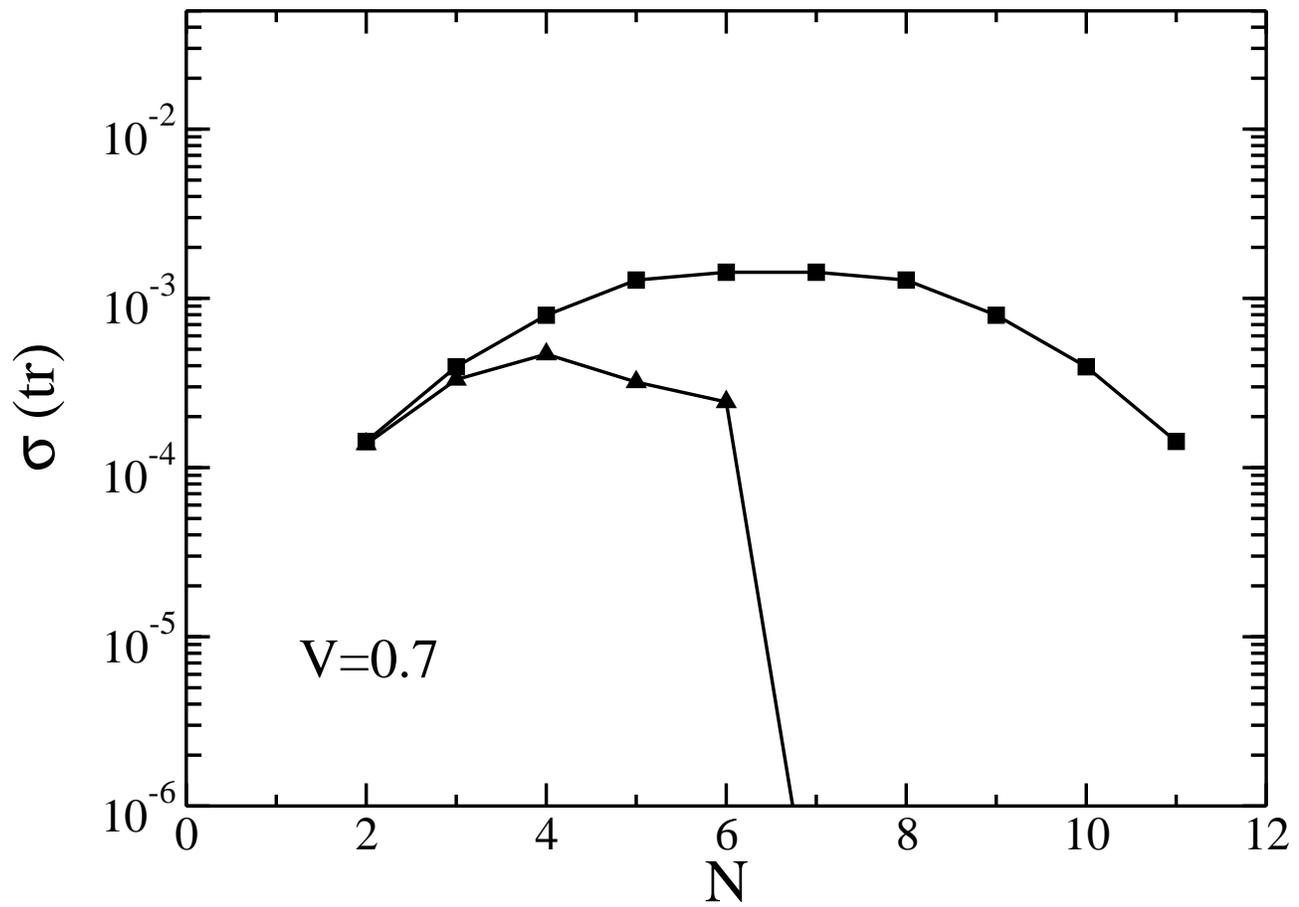}
\mbox{}\\[19cm]
\caption{As in Fig. 10, for $V=0.7$.}
\end{figure}

\end{document}